\documentstyle[sprocl,psfig,epsfig]{article}

\def\be{\begin{equation}}
\def\ee{\end{equation}}
\def\bea{\begin{eqnarray}}
\def\eea{\end{eqnarray}}

\begin{document}

\title{UNCERTAINTY OF THE SOLAR NEUTRINO ENERGY SPECTRUM
\footnote{Improved submitted version (April 15) to proceeding of 17
International Workshop on Weak Interactions and Neutrinos, South
Africa, January 24-30, 1999.}           
}

\author{Q. Y. Liu}

\address{The Abdus Salam International Center for
Theoretical Physics, \\ Strada Costiera 11,  34100, Trieste,
Italy\\E-mail: qiuyu@ictp.trieste.it}


\maketitle\abstracts{The solar neutrino spectrum measured by the
Super-Kamiokande shows an excess in high energy bins, which may be
explained by vacuum oscillation solution or $hep$ neutrino
effect. Here we reconsider an uncertainty of the data caused by the tail of
the energy resolution function. Events observed at energy higher than
$13.5$ MeV are  induced  by the tail of the resolution. At
Super-Kamiokande precision level this uncertainty is no more than few
percent within a Gaussian tail. But a 
 power-law decay tail at 3 $\sigma$  results considerable excesses
in these bins, which may be another possible
explanation of the anomaly in 708d(825d) data. }

\begin{figure}[t]
\mbox{\epsfig{figure=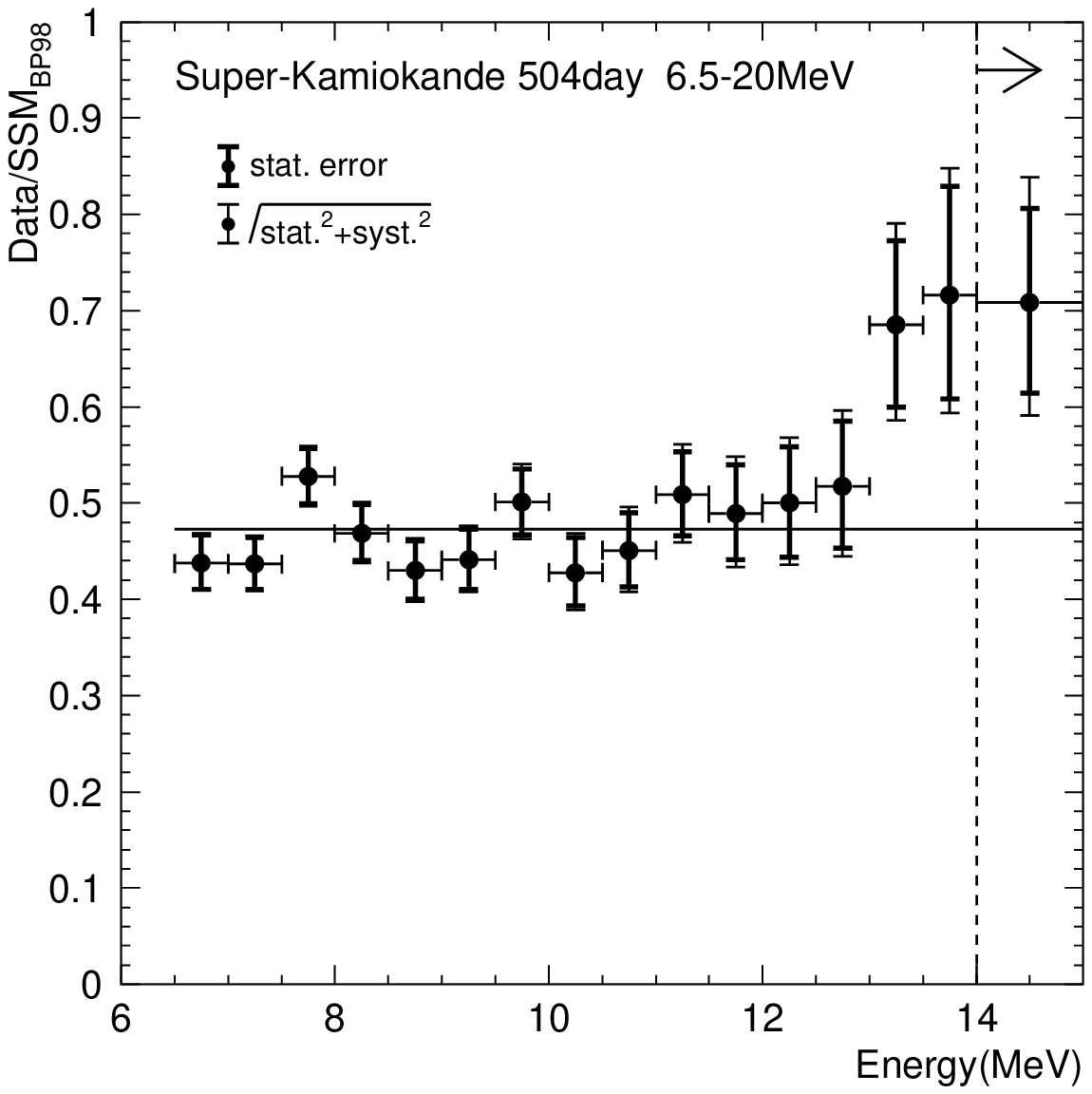,height=5.6cm}}
\mbox{\psfig{figure=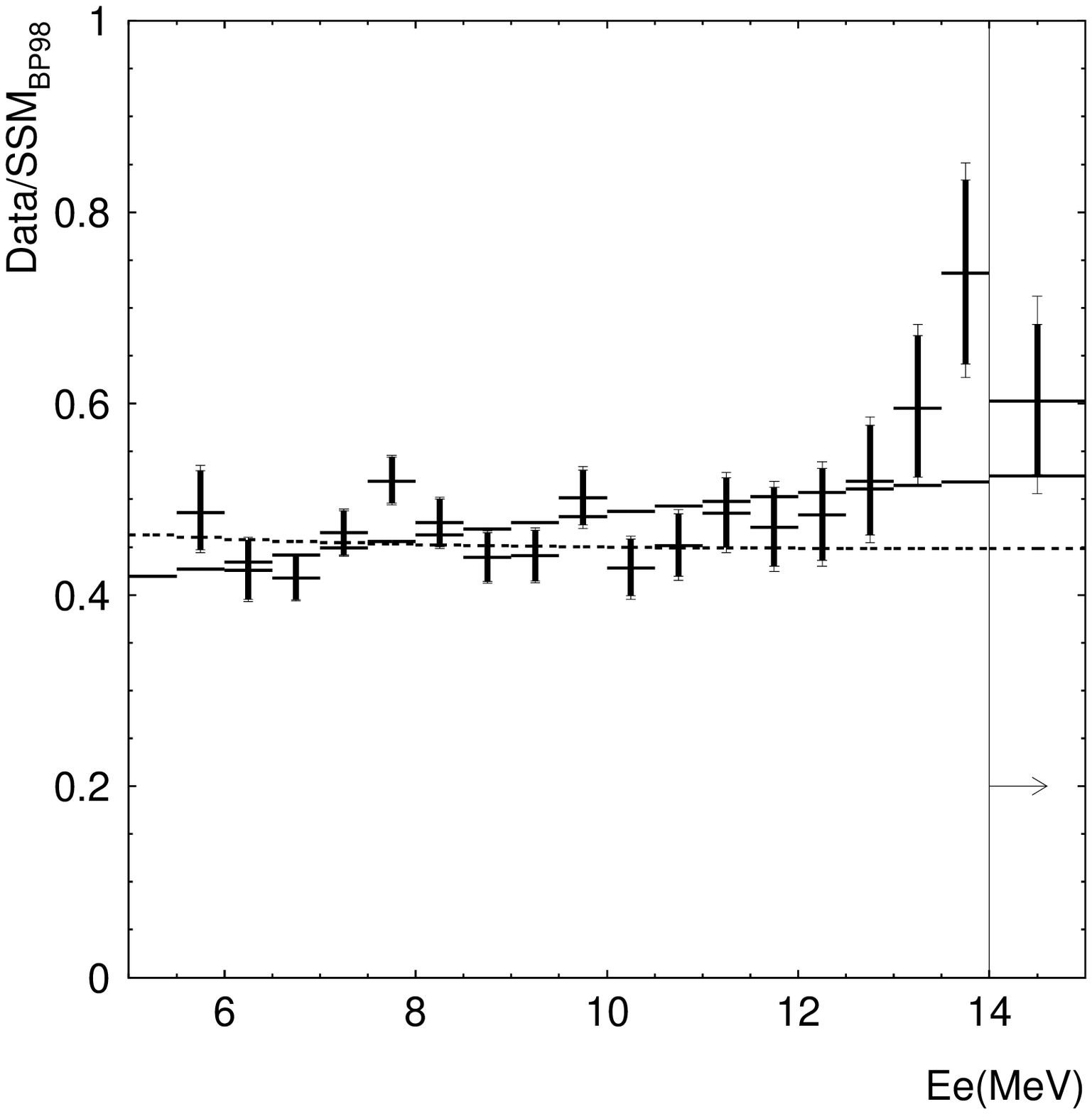,height=5.6cm}}
\caption{Ratio of observed electron energy spectrum and expectation
from the SSM. Left is for 504day, taken from reference [1]; right
is for 708day, histogram is MSW expectation, taken from reference [6].}
\label{fig:SPdata}
\end{figure}

A measurement of energy spectrum of recoil electrons from solar
neutrino scattering in the Super-Kamiokande detector shows excessive
events in high energy bins than what the usual oscillation solution
expects \cite{SK1}. The data was divided into 16 bins in energy scale,
every 0.5 MeV 
from 6.5 to 14.0 MeV and one bin combining events with energies from
14.0 to 20.0 MeV (fig. \ref{fig:SPdata}left). The 708 days data
has lower threshold down to 5.5 MeV (fig. \ref{fig:SPdata}right). This
excess may be due to a hep neutrino flux uncertainty, which requests a much
bigger cross section factor $S_{13}$ \cite{BahcallKrastev}; or it 
implies a vacuum oscillation solution to the solar neutrino
problem \cite{vacuumOsc}; it is also possible
that a fluctuation of the data or an absolute energy scale shifting
induce this excess \cite{SK1}.

Here we attempt to explain this anomaly by
the uncertainty in the tail of energy resolution
function of the detector \cite{liutalk}.            
The number of recoil electrons at a given real
energy $E_e$ equals:
\begin{equation}
\begin{array} {c}
T(E_e)~=~\int_{E_e - {m_e \over 2}} dE_{\nu} \cdot \Phi (E_{\nu})\cdot
~~~~~~~~~~~~
\\~~~~~~~~~~~~~~~~~~
{\displaystyle
\left[
P(E_{\nu}) {d \sigma_{\nu_e} \over
{dE_e }}(E_e,~E_{\nu})~ +
(1 - P(E_{\nu}))
{d \sigma_{\nu_{\mu}} \over {dE_e }}(E_e, ~E_{\nu})
\right]}
\end{array} 
\label{spectrumOrin}
\end{equation}
where $E_e$ is the total energy of recoil electron,
$\Phi (E_{\nu})$ is the original boron neutrino flux.
We can always use the oscillation survival probability
$P(E_{\nu})$ in (\ref{spectrumOrin}) since
$P(E_{\nu})=1$ stands for no neutrino 
conversion case. If a recoil electron has energy $E_e$,  the
water Cherenkov detector can not directly show us this 
real energy but another energy $E_{vis}$ (visible energy)
at a probability $f(E_{vis},~E_e)$. This is the energy resolution effect
of a detector and $f(E_{vis},~E_e)$ is energy resolution function. The
expected solar neutrino inducing electron spectrum is            
\begin{equation}
{\displaystyle S(E_{vis})~=~\int dE_e \cdot f(E_{vis},~E_e) T(E_e)}
\label{spectrum2}
\end{equation}

\begin{figure}[t]
\center{
\epsfig{figure=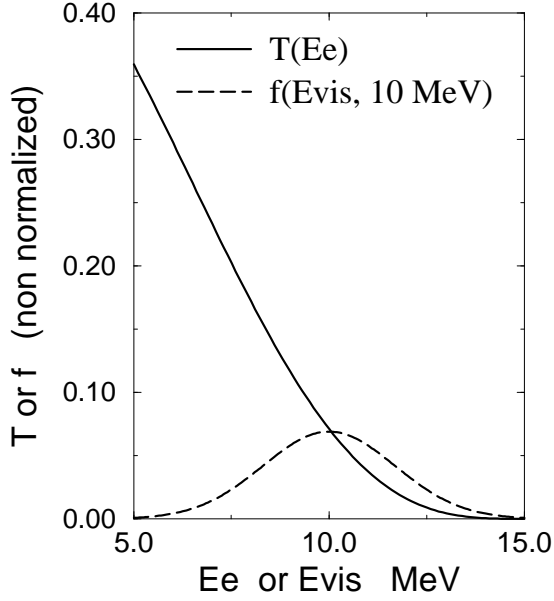,height=9cm}
}
\caption{The recoil electron spectrum (solid line) in real energy
$E_e$ scale and a standard energy resolution (dashed line) for
$E_e=10$ MeV in $E_{vis}$ scale.   \label{fig:Ee}} 
\end{figure}

The usual energy resolution function is parameterized as a
Gaussian function (fig. \ref{fig:Ee}) in the usual treatment:
\begin{equation}
{\displaystyle
f(E_{vis}, E_e)~=~
{1\over {\sqrt{2\pi} E_e \sigma(E_e)}}
\cdot
exp \left[
-\left( {E_{vis} - E_e \over {\sqrt{2}E_e\sigma (E_e)}}
\right)^2
\right ]
},
\label{resolutionf}
\end{equation}
where $\sigma \propto {1 \over \sqrt{E_e}}$.

This energy resolution function smears the spectrum in energy
scale. As fig. (\ref{fig:Ee}) shows, $T(E_e)$ is quite steep such that
the events above 13.5 MeV is negligible. In  $S(E_{vis})$ those
considerable events at this energy scale are 
totally  caused by a tail of the energy resolution function with lower
$E_e$ electrons. We then expect that the tail of the energy resolution
function in Super-Kamiokande experiment is relevant for the excess
events in above 13 MeV energy bins \cite{liutalk}. To integrate the
integrand in (\ref{spectrum2}) over $E_{vis}$ in one bin gives
\begin{equation}
I(E_e)~=~\int_{E_{vis}}^{E_{vis} + \Delta E_{bin}} dE_{vis} \cdot
f(E_{vis},~E_e) T(E_e) . 
\label{binspectrum}
\end{equation}
An important quantity $N_s(E_{e})$ which stands how many
$\sigma$ is the visible energy $E_{vis}$ far away from electron 
energy $E_e$, is introduced as
\begin{equation} 
N_s (E_e)~=~\frac{ |E_{vis} ~-~ E_e| }{ E_e \cdot \sigma (E_e) }
\label{Nsigma}
\end{equation}

\begin{figure}[t]
\mbox{\epsfig{figure=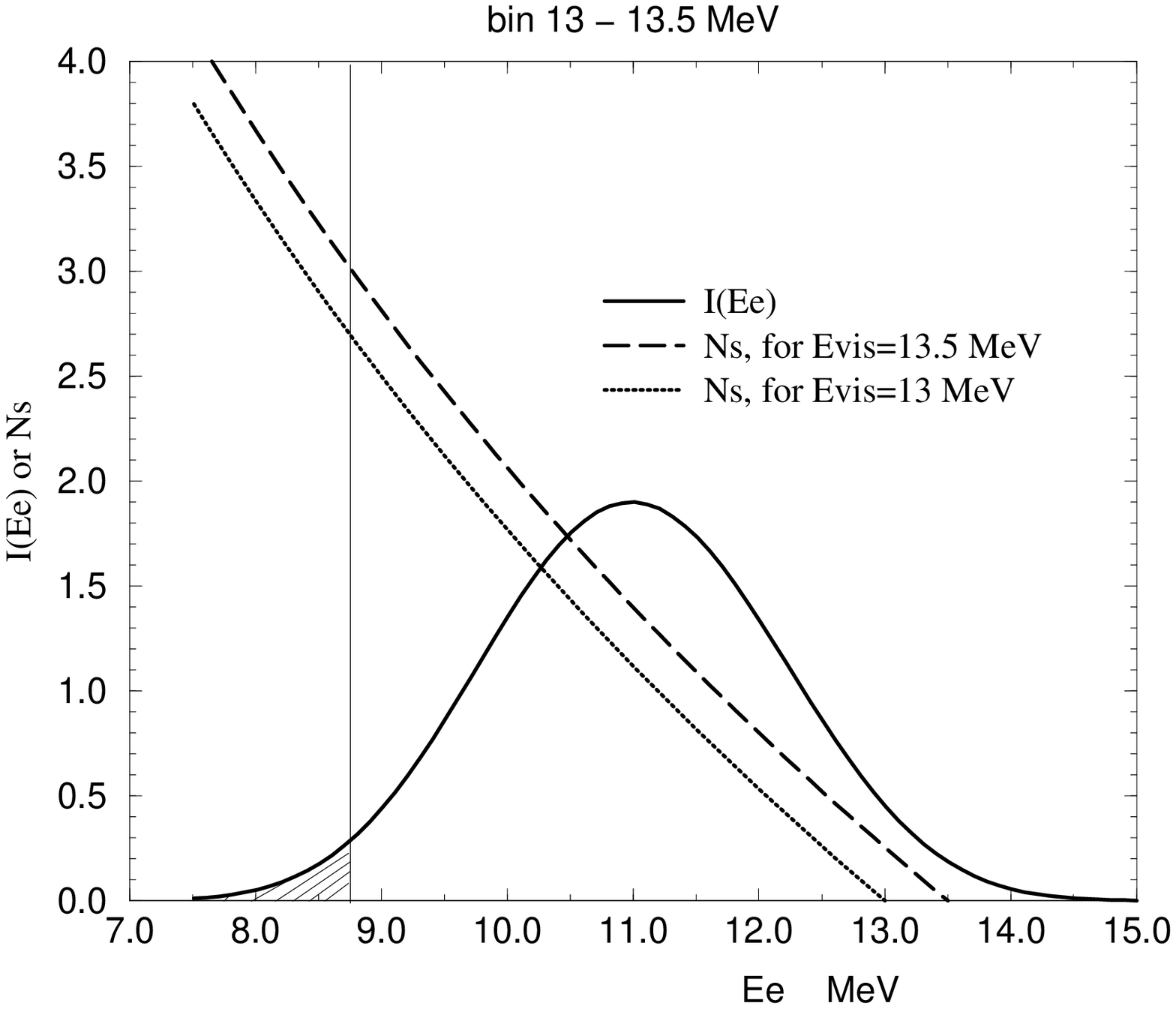,height=9cm}}
\mbox{\epsfig{figure=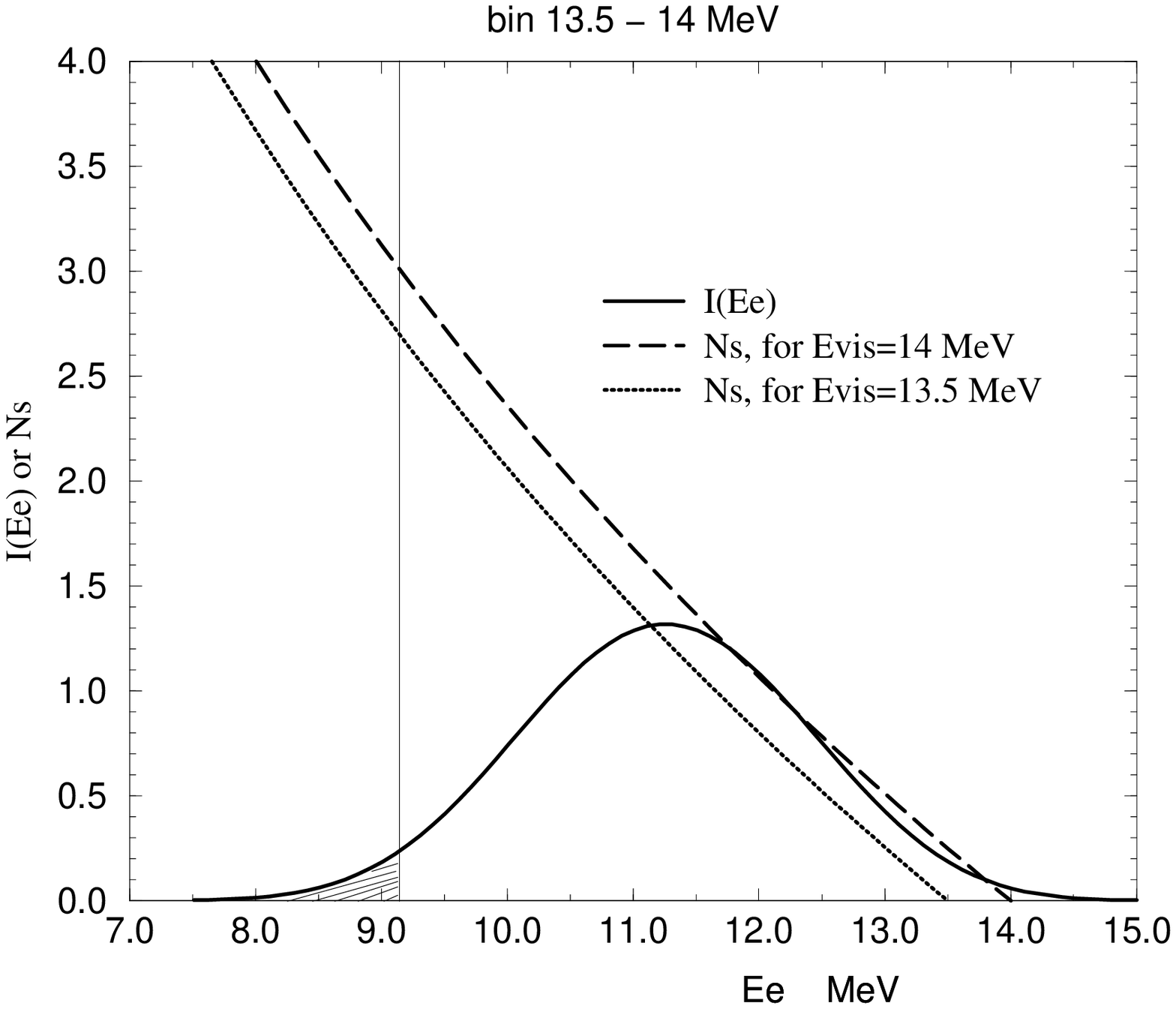,height=9cm}}
\caption{Contribution spectra to certain bins with respect to the real
electron energy $E_e$, see text.   \label{fig:binspectrum}}
\end{figure}

Fig. \ref{fig:binspectrum}up shows at a certain visible energy bin (13 -
13.5 MeV) the contribution spectrum $I(E_e)$ which is in solid
line and unnormalized. In the same plot the dotted(dashed) line is
$N_s(E_e)$ for $E_{vis}= 13(13.5)$ MeV, the unit of Y axis is one
standard deviation. $I(E_e)$ reaches its maximum at $11$ MeV, 
corresponding $N_s$ at 1.1 - 1.4 ($\sigma$). The
interval of $I(E_e)$ above the half of the maximum is $9.5$ - $12.4$
MeV, which corresponds to a $N_s$ range of 0.3 - 2.3 ($\sigma$). 
Fig. \ref{fig:binspectrum}down is for 13.5 - 14 MeV bin. In this bin
the maximum shifts to $E_e=11.25$ MeV while the above-half-height
corresponds $N_s$ from 0.5 to 2.5 ($\sigma$).

The Super-Kamiokande had calibrated their Monte Carlo energy
resolution function down to a tail of 3$\sigma$ by the LINAC
\cite{SK1}. The number of calibration events out of 4$\sigma$ tail is
just a few \cite{DPFWIN}. Thus an expected uncertainty comes from a
tail which is 3 $\sigma$ away from the center. In
fig. \ref{fig:binspectrum} the dashed area show how big is the
uncertainty part if we assume a standard Gaussian energy resolution.
It is no more than few percent. One need a much bigger resolution tail
to explain the 504d anomaly \cite{SK1,DPFWIN}.

However, the 708d data has simultaneously decreased two points in that three
bins by approximate one $\sigma$ than
what in the 504d data (fig. \ref{fig:SPdata}), 
This may re-open a possibility to explain the anomaly by experimental
uncertainties. 
Knowing that the detector's real energy resolution is not exact what shown
in (\ref{resolutionf}).
It can vary with respect to different points and different
trajectories. What we do is a kind of averaging.     
If the detector has a energy resolution tail which is
bigger than the tail of standard Gaussian, then it may cause this
excess in the high energy bins.

\begin{figure}[t]
\mbox{\epsfig{figure=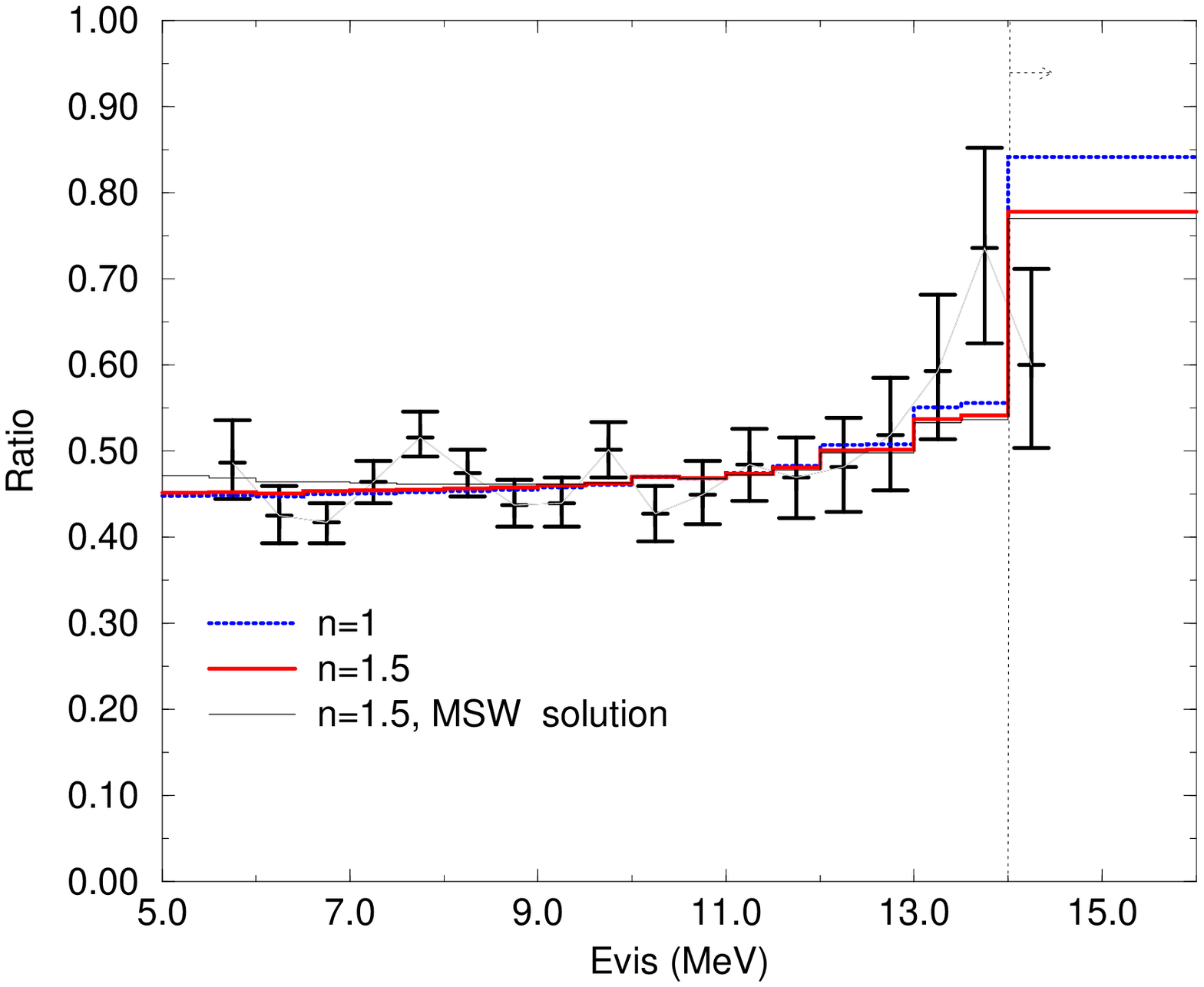,height=9cm}}
\caption{Expected spectra for UN-Gaussian tail out of 3 $\sigma$ with
$|1/(E_{vis} - E_e)|^n$ suppression, where the MSW parameters are
taken as
$\sin^2 2\theta = 0.79$ and $\Delta m^2 = 4 \cdot 10^{-5} ~eV^2$  .   \label{fig:3sigma.eps}}
\end{figure}

For example, a resolution function with a changed tail
\begin{equation}
{\displaystyle
f'(E_{vis}, E_e)~=~\left \{
\begin{array} {cc}
{1\over {\sqrt{2\pi} E_e \sigma(E_e)}}
\cdot
exp \left[
-\left( {E_{vis} - E_e \over {\sqrt{2}E_e\sigma (E_e)}}
\right)^2
\right ],
&
|E_{vis} - E_e| \leq 3 E_e \sigma(E_e)
\\ & \\
{1\over {\sqrt{2\pi} E_e \sigma(E_e)}}
\cdot
e^{-4.5}
\cdot |\frac{3 E_e \sigma(E_e)}{E_{vis} - E_e}|^n , 
&
|E_{vis} - E_e| \ge 3 E_e \sigma(E_e)
\end{array}  
\right .
}
\label{fprime}
\end{equation}
with a cut-off at 4.2 $\sigma$ results in the curves in
fig. \ref{fig:3sigma.eps}. Where $n$ is taken $1$ and $1.5$. 
In this plot we normalize in a way such
that a standard Gaussian tail will give a horizontal line at
Ratio=0.47. This distortion keeps almost steady when the solar
neutrino undergo large mixing MSW conversion \cite{MSW}.

Notice that
a spectrum by tight cut analysis shows that bins 13-13.5 MeV and
14-20 MeV have no more excess, only bin 13.5-14 MeV still has a ``bump''
\cite{DPFWIN} which can be fully due to a statistic fluctuation. This
spectrum is
also globally more flat than the regular cut spectrum. It may be
another supporting point that the anomaly is from experimental
uncertainty.

In conclusion, we have calculated the uncertainty part in high energy
bins of the solar neutrino energy spectrum. For a Gaussian energy
resolution it is no more than few percent. It seems this is not enough to
explain the excess of the 504d data in high energy bins. 
But it is still a possibility that the excess comes from experimental
uncertainty for the 708d data, as well as 825 days data.

\section*{Acknowledgments}
I am grateful to A.Yu. Smirnov and Y.L. Wu for useful discussions.

\end{document}